\documentclass[pdflatex,sn-mathphys-num]{sn-jnl}
\usepackage{graphicx} 
\usepackage{amsmath,amssymb,amsfonts}
\usepackage[nameinlink,capitalize,noabbrev]{cleveref}
\usepackage{braket}
\usepackage{algpseudocode}
\usepackage{algorithm}
\usepackage{qcircuit}
\usepackage{booktabs}
\usepackage{float}
\usepackage[table]{xcolor}
\usepackage{colortbl}

\definecolor{lightgray}{gray}{0.9}

\newcommand{\R}{\mathbb{R}} 
\newcommand{\ip}[1]{\langle #1 \rangle}

\begin{document}

\title[Triple-Hybrid quantum support vector classification]{A Triple-Hybrid Quantum Support Vector Machine Using Classical, Quantum Gate-based and Quantum Annealing-based Computing}

\author*[1]{\fnm{Juan C.} \sur{Boschero}}\email{juan.boschero@tno.nl}
\author*[1]{\fnm{Ward} \spfx{van der} \sur{Schoot}}\email{ward.vanderschoot@tno.nl}
\author*[1]{\fnm{Niels M. P.} \sur{Neumann}}\email{niels.neumann@tno.nl}

\affil*[1]{\orgdiv{Applied Cryptography \& Quantum Applications}, \orgname{The Netherlands Organisation of Applied Scientific Research (TNO)}, \orgaddress{\street{Anna van Buerenplein 1}, \postcode{2595DA} \city{The Hague}, \country{The Netherlands}}}

\abstract{Quantum machine learning is one of the fields where quantum computers are expected to bring advantages over classical methods. 
However, the limited size of current computers restricts the exploitation of the full potential of quantum machine learning methods. 
Additionally, different computing paradigms, both quantum and classical, each have their own strengths and weaknesses. 
Obtaining optimal results with algorithms thus requires algorithms to be tweaked to the underlying computational paradigm, and the tasks to be optimally distributed over the available computational resources. 
In this work, we explore the potential gains from combining different computing paradigms to solve the complex task of data classification for three different datasets. 
We use a gate-based quantum model to implement a quantum kernel and implement a complex feature map. 
Next, we formulate a quadratic unconstrained optimisation problem to be solved on quantum annealing hardware. 
We then evaluate the losses on classical hardware and reconfigure the model parameters accordingly. 
We tested this so-called triple-hybrid quantum support vector machine on various data sets, and find that it achieves higher precision than other support vector machines (both quantum and classical) on complex quantum data, whereas it achieves varying performance on simple classical data using limited training. 
For the complex data sets, the triple-hybrid version converges faster, requiring fewer circuit evaluations.}


\maketitle

\section{Introduction}
\subsection*{Background}
The quantum computing landscape is rapidly evolving, marked by the emergence of diverse quantum computation paradigms that promise to revolutionise computational capabilities across industries~\cite{marketsandmarkets2024quantum,McKinsey:2025}. 
Among these quantum paradigms, quantum annealers and gate-based quantum computers, together with classical systems, each offer unique strengths and limitations~\cite{salloum2024quantum,Neumann2023}. 
As the field matures, hybrid approaches that integrate these distinct technologies are gaining traction, particularly in machine learning applications where hybrid models can offer significant advantages~\cite{liu2022hybridgatebasedannealingquantum,Zardini2024}.

Various studies presented results on the potential of quantum machine learning. 
These results range from general overviews of quantum machine learning~\cite{Schuld:2015,Biamonte2017,Schuld2018} and specific quantum machine learning algorithms~\cite{Rebentrost2014,Schuld2019,Havl_ek_2019,vanDam2020,Meinhardt2020,Park2023,schnabel2024quantumkernelmethodsscrutiny,Xu_2024}, to hardware implementations of specific algorithms~\cite{Suzuki2024}.

Quantum machine learning combines quantum computing techniques with machine learning techniques to better solve problems. 
In general, quantum machine learning algorithms offer three different potential advantages~\cite{Neumann2019}:
Improved computation time, improved efficiency, higher capacities. 
The first advantage relates to the wall-clock time. 
Some operations, such as sampling complex probability distributions, can be performed more efficiently on quantum computers than on classical computers~\cite{Mansky:2023}. 
The second advantage relates to simpler models having the same predictive power as more complex classical models, or that more shorter/simpler training routines can result in the same overall performance~\cite{Liu2024}.
The third advantage relates to quantum associative networks having a higher storage capacity than classical alternatives~\cite{Rebentrost:2018,Meinhardt2020}. 
With this storage, these networks can retrieve the most likely patterns and correct errors.
Below we discuss the background on machine learning, kernels and hybrid computations. 

\subsection*{Support Vector Machines}
Classification is a common problem to tackle using quantum machine learning. 
These supervised machine learning algorithms use labelled data to train a classifier, which can then classify an unknown sample. 
Classically, support vector machines (SVMs) are often used for classification tasks. 
These SVMs try to find a separating hyperplane between two classes of a data set.
For separable data sets, this implies that there exist $\mathbf{w}\in\R^d$ and $b\in \R$ such that 
\begin{equation}
    y_i(\ip{\mathbf{w},\mathbf{x}_i} + b) \ge 1, \qquad \forall i=1,\hdots, M,
    \label{eq:hyperplane_equation}
\end{equation}
for data $\mathbf{x}_i$, labels $y_i$, and with $\ip{\cdot,\cdot}$ denoting the inner product function. 
For a new data point $\mathbf{x}$ we obtain its label by computing the sign of $\ip{\mathbf{w}, \mathbf{x}} + b$. 
For data that are not separable by a hyperplane, a looser formulation exists using slack variables. 

Several quantum approaches exist to implement SVMs. 
\citeauthor{Rebentrost2014} proposed a quantum algorithm that uses an efficient quantum routine for solving linear systems of equations~\cite{Rebentrost2014,HHL2009}. 
The algorithm by \citeauthor{Rebentrost2014} gives a sublinear-time procedure to classify a new data point. 
Later, other works also presented quantum approaches for SVMs on quantum annealing hardware~\cite{Willsch:2020} and on gate-based systems applied to indoor-outdoor detection~\cite{Phillipson:2021}.

\subsection*{Kernels}
Some data sets however do not admit a separation by hyperplane, even if some margin of error is allowed. 
Examples of such sets include radially separated data. 
The efficiency of the currently used models hinges on the fact that we can use such a separating hyperplane. 

By carefully manipulating the dataset, we can still use the efficient linear models on complex and non-linear data sets. 
This so-called kernel method is a common technique in the field of machine learning, where some map $\phi$ maps low-dimensional data to high-dimensional representations~\cite{Boser:1992}.
We call the higher-dimensional space the \textit{feature space} $\mathcal{F}$ and $\phi$ the \textit{feature map}. 
In the feature space, a non-linear non-trivial learning task becomes a linear and easy task. 
Every feature map gives rise to a kernel where a similarity value between two data points $\boldsymbol{x}$ and $\boldsymbol{x'}$ can be obtained~\cite{schnabel2024quantumkernelmethodsscrutiny}. 

Note that evaluating the feature map for all data points can be costly and might be necessary for specific machine learning algorithms. 
However, for our purpose, it suffices to only compute the inner product between the feature space representation of our data points.

Note that for radially separated data a kernel can map a data point to a higher representation, where one variable depends on the norm of the original data point. 
A feature map that exploits this feature would then help to obtain a linear separation between the two classes in the feature space. 

\subsection*{Quantum kernels}
Various works in literature have tried to solve SVMs or increase their performance using quantum methods~\cite{Schuld2019,Havl_ek_2019,Park2023,schnabel2024quantumkernelmethodsscrutiny,Xu_2024}.
A quantum kernel $K$ can be written as $K(\mathbf{x},\mathbf{z}) = | \braket{\Phi(\mathbf{x}) | \Phi(\mathbf{z})} |^2$. 
The choice of feature map $\phi$ determines the potential of the kernel to a large extent. 
If the chosen feature map is too simple, the quantum kernel will have no advantage over classical kernels, whereas a too complex quantum kernel will be expensive to evaluate.

Ref.~\cite{Havl_ek_2019} proposed a specific feature map for an $n$-dimensional vector $\mathbf{x}$ given by 
\begin{equation}
    \ket{\Phi(\mathbf{x})} = U_{\phi(\mathbf{x})}H^{\otimes n}U_{\phi(\mathbf{x})}H^{\otimes n} \ket{0^{\otimes n}},
    \label{eq:feature_map}
\end{equation}
where $H$ is a Hadamard gate and,
\begin{equation}
    U_{\phi(\mathbf{x})} = \exp\bigg(i\sum_{S\subseteq [n]} \phi_S(\mathbf{x}) \prod_{i \in S } Z_i \bigg).
    \label{eq:feature_map_operator}
\end{equation}
Here, $\phi_S$ is a map applied to some subset $S\subseteq [n]$. 
Note that as the dimension of the data grows, so does the complexity of implementing the feature map. 
Therefore, the set $S$ is often restricted to a size of at most $2$. 
As an additional benefit, this restriction assures that the feature map has Ising-like properties, which makes them suitable for quadratic optimisation. 
This specific feature map is often referred to as the $ZZ$-feature map, as at most two terms interact with each other and $Z$ gates govern the interactions. 

\subsection*{Hybrid computations}
Quantum computers thus have great potential in the area of machine learning, especially for complex data sets. 
However, no single quantum paradigm currently dominates across all problem domains. 
Instead, each system excels in specific computational tasks: 
quantum annealers, such as those developed by D-Wave, are suited for solving combinatorial optimisation problems through energy minimisation~\cite{cohen2014}, while gate-based quantum computers, like those from IBM and Rigetti~\cite{alexandrou2021,quantum2020019,AbuGhanem2025}, offer universal quantum computation with high flexibility and precision. 
At the same time, classical computers remain indispensable for data preprocessing, orchestration, and post-processing tasks.

The fundamental differences between quantum annealers and gate-based systems are rooted in their operational principles. 
Quantum annealers rely on adiabatic evolution to find low-energy states of an Ising model, making them well-suited for optimisation problems but limited in general-purpose quantum algorithms~\cite{KadowakiNishimori:1998}. 
They can be programmed by presenting a problem in a quadratic unconstrained optimisation (QUBO) formulation. 
In contrast, gate-based quantum computers manipulate qubits through sequences of quantum gates, enabling the implementation of a wide range of algorithms, including those for linear algebra and quantum machine learning~\cite{Nielsen_Chuang_2010}.
They are programmed via gates manipulating quantum states.
These differences make them ideal candidates for a hybrid architecture, where each system contributes according to its strengths.

Recent developments have shown growing interest in hybrid computations, combining different computational backends in a single workflow~\cite{Phillipson2023}. 
The work presented in~\cite{Jattana_2024} introduces a foundational triple-hybrid method, where a quantum annealer is employed to optimise the variational parameters of a Variational Quantum Eigensolver (VQE) circuit, aiming to determine the ground state energy of a hydrogen molecule ($\mathrm{H}_2$).
Another work explores a triple-hybrid framework for drone routing problems~\cite{osaba2025solvingdroneroutingproblems}. 
In that approach, a gate-based quantum computer solves a formulated Max-Cut problem, and the resulting solutions are evaluated by a quantum annealer to select the best route from the available options.
Next, other methods combine two different quantum systems or a quantum and a classical system to solve problems.
Examples of such two-hybrid systems include non-local distributed computations~\cite{Caleffi:2018,neumann2020imperfect,Caleffi:2024,Boschero:2025} and variational algorithms~\cite{Peruzzo2014,Cerezo2021}.
This paper introduces a novel triple-hybrid quantum support vector machine (QSVM) framework that leverages the complementary capabilities of a quantum annealer, a gate-based quantum computer, and a classical computer. 

\subsection*{Contribution and structure of the paper}
This research presents an implementation of a triple-hybrid approach and applies it to the task of classification. 
We revisited an already existing approach and made a new implementation where various computational platforms were combined. 
In this triple-hybrid approach, the classical kernel traditionally used to solve optimisation problems is replaced by a quantum kernel to address more complex classification tasks. 
In particular, classification is a broadly applicable problem, making this approach versatile across a wide range of domains.
By integrating three computational paradigms: quantum annealing for optimisation, gate-based quantum computing for algorithmic flexibility, and classical computing for stability and scalability, the proposed quantum support vector machine (QSVM) framework aims to enhance classification performance. 
Moreover, this hybrid strategy offers a scalable and forward-looking solution as quantum hardware continues to advance.

The remainder of this paper starts with the mathematical foundation of our work. 
This section outlines how optimisation-based quantum systems can be adapted for classification tasks.
This section builds upon existing literature and aims to make the work self-contained.
In the experimental section, we describe the data set used and detail the methodology for embedding a quantum circuit-derived kernel into a QUBO formulation, which is then processed by a quantum annealer for classification. 
The results section presents a comparative analysis of three approaches: a purely classical SVM, a quantum gate-based SVM, and our proposed triple-hybrid method. 
We compare three data sets: two simple classical data sets, and a hard data set based on quantum data. 
Finally, we discuss the implications of our findings and conclude with insights into the potential and limitations of the triple-hybrid method. 

\section{Mathematical background}
This section gives mathematical support for the SVM implementation presented in this work. 
We follow the definitions used in~\cite{Havl_ek_2019,Willsch:2020} in defining our SVM. 

\subsection{Classification using SVMs}
Given a dataset consisting of $M$ pairs $(\mathbf{x}_i, y_i)\in\R^d\times\{-1,1\}$, with data points $\mathbf{x}_i$ and labels $y_i$. 
You are promised there exists some function $f:\R^d\to \{-1,1\}$ such that $f(\mathbf{x}_i)=y_i$ for every $i=1,\hdots,M$. 
The goal of a classifier is to predict $f(\mathbf{x})$ for some new unseen $\mathbf{x}$. 

Naturally, for arbitrary $f$, this problem cannot be solved with probability $1/2+\varepsilon$ for any $\varepsilon>0$, as no information is known about $f$. 
However, practical data sets typically have some internal structure, implying also that $f$ has some internal structure. 

For now assume that the data set is separable by some hyperplane. 
We saw already that we can take Equation~\eqref{eq:hyperplane_equation} and formulate an optimisation problem from it. 
However, simply minimising the norm of $\mathbf{w}$ will give a zero outcome, and the classifier will no longer work. 
To assure that the constraints remain satisfied, Lagrange multipliers $\alpha_i\ge 0$ are introduced, giving the optimisation problem
\begin{equation}
    \min_{\mathbf{w},b,\alpha_i} \frac{||\mathbf{w}||_2^2}{2} - \sum_{i=1}^{M} \alpha_i y_i(\ip{\mathbf{w},\mathbf{x}_i} + b) + \sum_{i=1}^{M} \alpha_i.
    \label{eq:optimisation_problem}
\end{equation}

For dataset not separable by a hyperplane, we introduce a margin of error using slack variables $\xi_i \ge 0$.
Equation~\eqref{eq:hyperplane_equation} then becomes 
\begin{align}
    y_i(\ip{\mathbf{w},\mathbf{x}_i} + b) \ge 1-\xi_i, \qquad \forall i=1,\hdots, M. \label{eq:hyperplane_equation_slack}
\end{align}
Similarly, the optimisation problem is then given by
\begin{equation}
    \min_{\mathbf{w},b,\alpha_i,\xi_i, \mu_i} \frac{||\mathbf{w}||_2^2}{2} + C\sum_{i=1}^{M} \xi_i + \sum_{i=1}^{M} \mu_i \xi_i - \sum_{i=1}^{M} \alpha_i y_i(\ip{\mathbf{w},\mathbf{x}_i} + b) + \sum_{i=1}^{M} \alpha_i,
    \label{eq:optimisation_problem_slack}
\end{equation}
for some constant $C>0$. 

The Lagrange formulation of the optimisation problem has the advantage that it allows us to consider the dual formulation of the problem. 
Formulating the problem in terms of the dual problem often leads to an easier to solve optimisation problem.

\subsection{Dual version}
The dual formulation of the problem follows from taking the partial derivatives of the Lagrange formulation with respect to $b$, $\mathbf{w}_j$ and $\mu_j$. 
The Karush-Kuhn-Tucker conditions then guarantee that an optimal solution to the original and dual version of the problem, should it exist, has to satisfy regularity constraints~\cite{Karush:1939,KuhnTucker:1951}.
For the partial derivatives we find
\begin{align*}
    \frac{\partial}{\partial b} & = -\sum_{i=1}^{M} \alpha_i y_i = 0, \\
    \frac{\partial}{\partial \mathbf{w}_j} & = \mathbf{w}_j - \sum_{i=1}^{M} \alpha_i y_i \mathbf{x_i}_j = 0, \\
    \frac{\partial}{\partial \mu_j} & = \xi_i = 0.
\end{align*}
The partial derivatives with respect to $\mathbf{w}_j$ yield
\begin{equation}
    \mathbf{w} = \sum_{i=1}^{M} \alpha_i y_i \mathbf{x_i}.
\end{equation}
Noting that $||\mathbf{w}||_2^2 = \mathbf{w}^{T} \mathbf{w}$, we now use this expression for $\mathbf{w}$ and substitute it into Equation~\eqref{eq:optimisation_problem_slack} to obtain
\begin{equation}
    \min_{\mathbf{w},b,\alpha_i} \sum_{i=1}^{M} \alpha_i - \frac{1}{2}\mathbf{\alpha y}^{T} \cdot \mathbf{K} \cdot \mathbf{\alpha y}, \label{eq:optimisation_problem_dual}
\end{equation}
where $\mathbf{\alpha y}$ is the element-wise product of the vectors $\mathbf{\alpha}$ and $\mathbf{y}$ and $\mathbf{K}$ is a matrix such that $\mathbf{K}_{ij} = \ip{\mathbf{x_i},\mathbf{x_j}}$. 
The corresponding prediction function $f$ for a new data point $\mathbf{x}$ now consists of computing the sign of $\sum_{i=1}^{M} \alpha_i y_i \ip{\mathbf{x_i}, \mathbf{x}}$.
From Equation~\eqref{eq:optimisation_problem_dual}, we can obtain a QUBO formulation, which we will use in the experiments later on.
The QUBO formulation for Equation~\eqref{eq:optimisation_problem_dual} is given by 
\begin{equation}
    \frac{1}{2}\mathbf{y}^{T} \cdot \mathbf{K} \cdot \mathbf{y}.
    \label{eq:formulation_QUBO}
\end{equation}

This matrix $K$ also extends to kernels. 
In this work, we will also apply a kernel to our data, and we use a gate-based quantum computing approach to fill in the kernel. 
Specifically, we obtain three different variants for the hybrid SVM: 
1) A fully classical SVM;
2) A quantum SVM with a quantum kernel solved using classical methods;
and, 3) A quantum SVM with a quantum kernel solved using quantum annealing.
The third setup yields a triple-hybrid algorithm consisting of classical, quantum gate-based, and quantum annealing operations.
As far as the authors are aware, this work is the first that considers a triple-hybrid approach to the SVM problem. 
In the next sections, we compare these setups.
    
\section{Experimental setup}
This section describes the three data sets used in the experiments, as well as the method used to carry out the experiments. 

\subsection{Used datasets}
This work uses three different data sets. 
The first data set, the breast cancer data set~\cite{breast_cancer_wisconsin_diagnostic_17}, has 30 features and a single target (malignant or benign).
Out of the 30 features, ten real-valued features are computed for each cell nucleus. 
For simplicity, the radius (mean of distances from the centre point to the other points in the perimeter) and the texture (standard deviation of gray-scale values) were selected as the chosen features in the analysis.

The second banknote authentication data set~\cite{banknote_authentication_267} has four features and a single target (authenticated or not). The two selected features are the variance and the skewness of the Wavelet Transformed image.

Both data sets contain classical data. 
The third data set was artificially generated, to contain characteristics typically hard for classical methods. 
As such, we expect that the performance on the third data set will be to some degree representative for the performance on other data sets that classical methods struggle with. 
Specifically, we used the \textsc{ad\_hoc\_data} method by Qiskit to generate the data~\cite{javadiabhari2024quantumcomputingqiskit}.
This method generates a data set that can be fully separated by the $ZZ$-feature map shown in Equations~\eqref{eq:feature_map} and~\eqref{eq:feature_map_operator}.

This method first samples a uniformly distributed random vector $\mathbf{x}\in(0,2\pi]^n$ and then applies the feature map from Equation~\eqref{eq:feature_map}. 
The labels of the data set are now distributed according to 
\begin{equation}
   f(\vec{x}) = \begin{cases} 
      1 & \braket{\Phi(\vec{x})|V^\dagger \prod_i Z_iV| \Phi(\vec{x}) } > \Delta \\
      -1 & \braket{\Phi(\vec{x})|V^\dagger \prod_i Z_iV| \Phi(\vec{x}) } < -\Delta  
   \end{cases}
\end{equation}
where $\Delta$ is a separation gap and $V \in \mathrm{SU(4)}$ is a random unitary~\cite{Havl_ek_2019}. 
We used $n=2$ for the experiments in this paper. 

The Qiskit method guarantees that the data set is separated with gap $\Delta$. 
With data sets without such a guaranteed gap, the labels can still be assigned via a similar rule, where labels that are within some predefined gap are assigned randomly. 
This assignment can be uniformly at random or weighted based on the magnitude of the inner product. 
In this work, this data generation method is used with an arbitrary gap of $\Delta = 0.6$. 
For reproducibility, we fixed the data generation method in our experiment with seeds 300, 600 and 1000~\cite{javadiabhari2024quantumcomputingqiskit}.

\subsection{Method}
The methodology consists of four main steps. 
First, a certain each data set, a training data set is generated of varying size (50, 100, 200, 300 and 500 data points). Next, a test set consisting of five times fewer validation points are selected uniformly at random. 
\begin{figure}[b]
\centering
    \centerline{\includegraphics[width=1\textwidth]{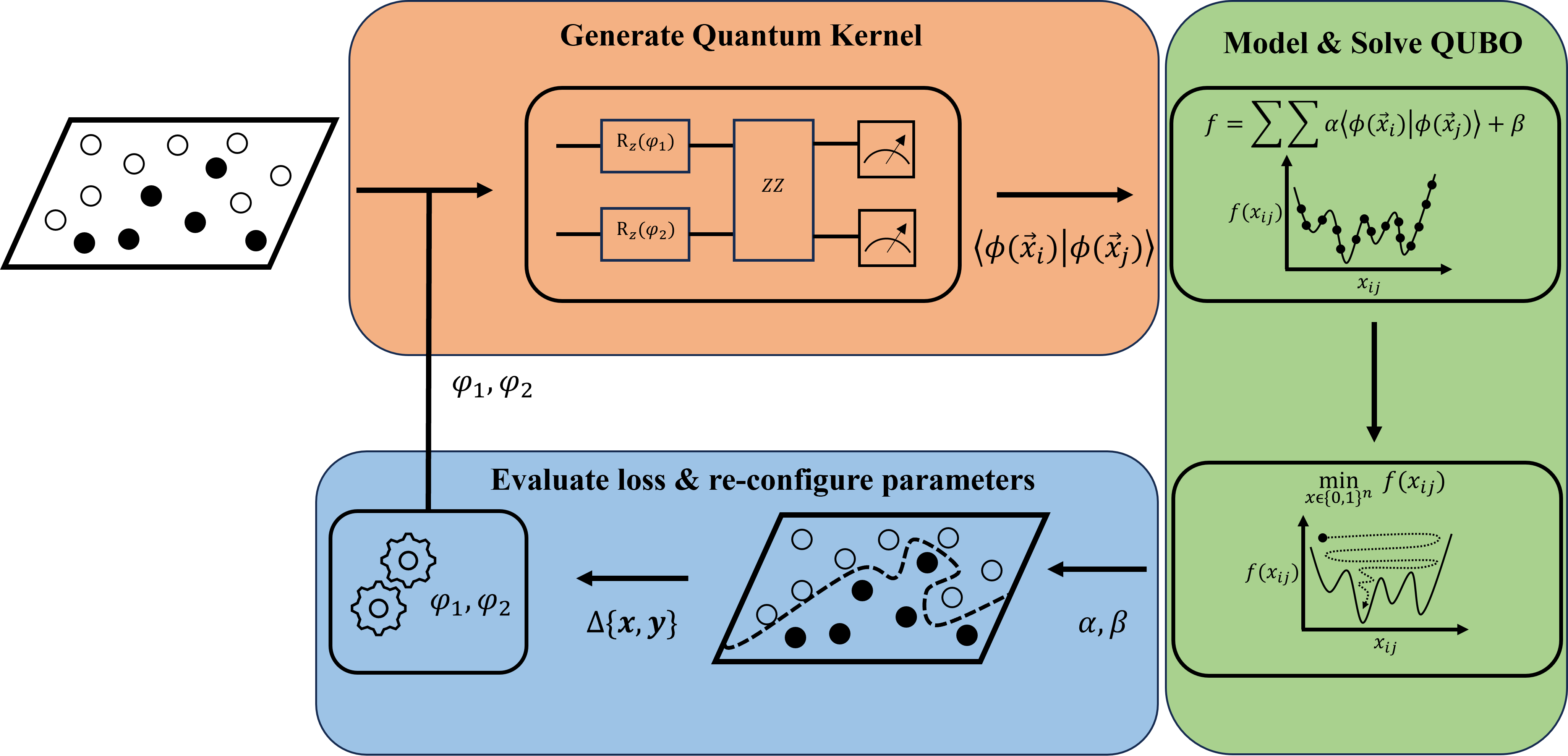}}
    \caption{The quantum hybrid process begins by breaking down the data into essential components needed to create a quantum circuit for the kernel. 
    After generating the circuit, a QUBO function is prepared by sampling the quantum kernel. 
    This QUBO is then solved and the outcomes are used to compute the $\alpha$ and $\beta$ values used in the classification function. 
    The loss function is evaluated and fed into an optimiser, which selects the training parameters for the quantum kernel. 
    This cycle repeats until the desired accuracy is achieved or the maximum number of iterations is reached. 
    The orange box depicts the processes executed by a gate-based quantum computer, the green box denotes the processes executed by a quantum annealer and the blue box shows the processes carried out by a classical computer.}
    \label{fig:process}
\end{figure}
Next, the quantum kernel is generated. 
The quantum kernel is generated by preparing a second-order Pauli-$Z$ evolution circuit parametrised using randomly selected training parameters, $\varphi_n \in [-2\pi, 2\pi]$.
The data set determines the number of training parameters. 
The experiments shown in this work use two training parameters $\varphi_1$ and $\varphi_2$. 
Although the data are subsequently parsed through the circuit, there is no initial pre-processing done to estimate the value of the training parameters.

Third, the QUBO is modelled and solved. 
The algorithm uses the training data and training labels to construct the QUBO, following Equation~\eqref{eq:formulation_QUBO}.
After modelling the QUBO, it is solved using a (simulated) quantum annealer.
The outputs, $\alpha$ and~$\beta$, are subsequently used in the classification function to estimate the label of a given point.
The computational formulation is detailed in Algorithm~\ref{alg:solve_qubo}.

The final step is to fit the test data using the classification function.
This allows for the calculation of the model's total accuracy.
An optimiser uses this total accuracy to update the training parameters and generate a new quantum kernel.
The optimiser used for this process is \textit{Constrained Optimization by Linear Approximation} (COBYLA), an iterative method for derivative-free constrained optimization~\cite{conn2009introduction}. 
COBYLA constructs and updates linear approximations of both the objective function and each constraint, making it more efficient than gradient-based methods such as L-BFGS. 
If the accuracy exceeds a user-defined threshold, the process terminates. 
Otherwise, it iterates through steps two to four until the desired accuracy is achieved. 
Figure~\ref{fig:process} graphically outlines the four steps, where each box denotes a different computing paradigm.

All gate-based quantum operations are simulated using Qiskit's Aer simulator and the quantum annealing results are obtained using simulated quantum annealing offered by D-Wave.
The classical computations are run on a standard laptop. 

\begin{algorithm}
\caption{Solve QUBO}\label{alg:solve_qubo}
\begin{algorithmic}[1]
\Require Training data $\mathbf{X^{data}}$, Training labels $\mathbf{X^{labels}}$ 
\State $N \gets$ size of $\mathbf{X^{data}}$
\State Initialize $Q$ as a zero matrix of size $N^2$
\For{$n = 0$ to $N-1$}
    \For{$m = 0$ to $N-1$}
        \State $kernel_{tot} \gets X^{label}_n X^{label}_m$
        \If{$m \neq n$}
            \State $k_{val} \gets |\braket{\Phi(X^{data}_{m}) | \Phi(X^{data}_{n})}|^2$
            \State $kernel_{tot} \gets kernel_{tot} + k_{val}$
            \State $Q[m, n] \gets -\frac{1}{2} kernel_{tot} X^{labels}_m X^{labels}_n$
        \EndIf
    \EndFor
\EndFor

\State $q_{size} \gets$ size of $Q$
\State $qubo \gets \{(i, j) : Q[i, j] \mid i, j \in [0, q_{size}-1] \}$

\State $\alpha \gets $ sample $qubo$ using annealer

\State $\beta \gets \frac{1}{N} \sum\limits_{n=0}^{N-1} \Big(X^{labels}_n - \sum\limits_{m=0}^{N-1} \alpha_m X^{labels}_m |\braket{\Phi(X^{data}_{m}) | \Phi(X^{data}_{n})}|^2 \Big)$

\State \Return $\alpha, \beta$
\end{algorithmic}
\end{algorithm}

\section{Results}
In this section, the results of the experiments are detailed. The results will be discussed per data set. For each solver and data set, a maximal of 10 iterations was used, hence, results might be better with more training iterations. 
We consider a setting with limited resources, thus limiting the number of training iterations. 

\subsection{Breast cancer and Bank datasets}
The results are presented in Table \ref{table:results_Bank} and \ref{table:results_BC}. A maximum of 10 iterations were used across all three methods and the iteration with the highest accuracy was ultimately chosen. 

For the bank data set, we observe that, except for the smallest data set, the classical SVM has difficulty achieving good accuracy. 
Especially for the larger data sets the HQSVM performs slightly better than its classical counterpart. 
Still all three solvers fail to train a high-performing classifier within the given training constraints. 
Averaging all scores per method for the different data sets, the classical SVM achieved 60.73\% accuracy, the QSVM 52.24\%, and the HQSVM 64.67\%.
The HQSVM thus attained the highest overall accuracy within the allowed number of training iterations, though using more training iterations than the classical method. 

For the Breast cancer data set, the picture is very different. Firstly, it can be seen that the classical and QSVM methods perform quite well, with the classical solver slightly outperforming the QSVM method (average accuracy of 90.86\% compared to 88.31\%). Although the HQSVM solver did not perform as poorly on average as it did in the bank dataset, it still attained the lowest average of 80.27\% compared to the other two methods. 

\begin{table}
\centering
\caption{Experiment results across different training/testing points and seeds for the Bank dataset.}
\label{table:results_Bank}
\scriptsize
\setlength{\tabcolsep}{5pt} 
\renewcommand{\arraystretch}{1.1} 

\begin{tabular}{|c|c|ccc|ccc|}
\hline
\textbf{Train} & \textbf{Test} & 
\multicolumn{3}{c|}{\textbf{Accuracy (\%)}} & 
\multicolumn{3}{c|}{\textbf{Iterations}} \\
\textbf{pts} & \textbf{pts} & 
\textbf{Classical} & \textbf{QSVM} & \textbf{HQSVM} &
\textbf{Classical} & \textbf{QSVM} & \textbf{HQSVM} \\
\hline
\rowcolor{lightgray}
50  & 10    &  90.00  &  61.90   &  50.00   &  6 & 10  & 6 \\
100 & 20    &  50.00  &  58.33   &  90.00   &  4 &  10 & 8 \\
\rowcolor{lightgray}
200 & 40    &  50.00  &   57.50  &   50.00  & 3 & 8  & 10\\
300 & 60    &  56.67  & 33.48  & 68.34  & 3  & 10  &  6\\
\rowcolor{lightgray}
500 & 100 & 57.00 & 50.00 & 65.00 & 3 & 3 & 8 \\
\hline
\end{tabular}
\end{table}

\begin{table}
\centering
\caption{Experiment results across different training/testing points and seeds for the Breast cancer dataset.}
\label{table:results_BC}
\scriptsize
\setlength{\tabcolsep}{5pt} 
\renewcommand{\arraystretch}{1.1} 

\begin{tabular}{|c|c|ccc|ccc|}
\hline
\textbf{Train} & \textbf{Test} & 
\multicolumn{3}{c|}{\textbf{Accuracy (\%)}} & 
\multicolumn{3}{c|}{\textbf{Iterations}} \\
\textbf{pts} & \textbf{pts} & 
\textbf{Classical} & \textbf{QSVM} & \textbf{HQSVM} &
\textbf{Classical} & \textbf{QSVM} & \textbf{HQSVM} \\
\hline
\rowcolor{lightgray}
50  & 10    & 90   &   100  &  83.34   & 6  &  6 & 8 \\
100 & 20    &  90  &  81.87   &  76.33   &  6 & 3  &  3 \\
\rowcolor{lightgray}
200 & 40    &  90  &  88   &  83.34   & 6  & 6  & 10\\
300 & 60    &  91.67  &  80.50 &  83.34 &  8 &  5 & 10 \\
\rowcolor{lightgray}
500 & 100 & 92.63 & 91.19 & 75.00 & 8 & 3 & 2\\
\hline
\end{tabular}
\end{table}

\subsection{Qiskit dataset}
For the quantum data set, we used multiple seeds to generate the data, thereby obtaining multiple data sets. 
Notably, due to the data's bias towards quantum kernels, purely classical techniques yielded highly erroneous results.
Therefore, a classical logistic regression was fitted on a simulated quantum kernel to achieve more accurate outcomes.

\begin{table}
\centering
\caption{Experiment results across different training/testing points and seeds for the Qiskit dataset.
Note that the classical results include an additional fitted logistic regression to mitigate the errors.}
\label{table:results}
\scriptsize
\setlength{\tabcolsep}{5pt} 
\renewcommand{\arraystretch}{1.1} 

\begin{tabular}{|c|c|c|ccc|ccc|}
\hline
\textbf{Train} & \textbf{Test} & \textbf{Seed} & 
\multicolumn{3}{c|}{\textbf{Accuracy (\%)}} & 
\multicolumn{3}{c|}{\textbf{Iterations}} \\
\textbf{pts} & \textbf{pts} & \textbf{num} & 
\textbf{Classical} & \textbf{QSVM} & \textbf{HQSVM} &
\textbf{Classical} & \textbf{QSVM} & \textbf{HQSVM} \\
\hline
\rowcolor{lightgray}
50  & 10  & 300  & 100   & 100    & 100    & 6  & 8  & 1 \\
\rowcolor{lightgray}
50  & 10  & 600  & 94    & 100    & 100    & 6  & 8  & 1 \\
\rowcolor{lightgray}
50  & 10  & 1000 & 100   & 100    & 100    & 6  & 8  & 1 \\
100 & 20  & 300  & 100   & 100    & 100    & 8  & 4  & 1 \\
100 & 20  & 600  & 97.5  & 100    & 100    & 6  & 3  & 4 \\
100 & 20  & 1000 & 100   & 100    & 100    & 7  & 8  & 4 \\
\rowcolor{lightgray}
200 & 40  & 300  & 100   & 100    & 100    & 10 & 5  & 1 \\
\rowcolor{lightgray}
200 & 40  & 600  & 100   & 100    & 100    & 6  & 10 & 4 \\
\rowcolor{lightgray}
200 & 40  & 1000 & 100   & 100    & 100    & 8  & 10 & 4 \\
300 & 60  & 300  & 100   & 93.33  & 99.17  & 8  & 7  & 2 \\
300 & 60  & 600  & 100   & 93.33  & 99.17  & 9  & 2  & 4 \\
300 & 60  & 1000 & 100   & 96.67  & 99.17  & 8  & 2  & 6 \\
\hline
\end{tabular}
\end{table}

\begin{figure}[b]
\centering
    \centerline{\includegraphics[width=0.8\textwidth]{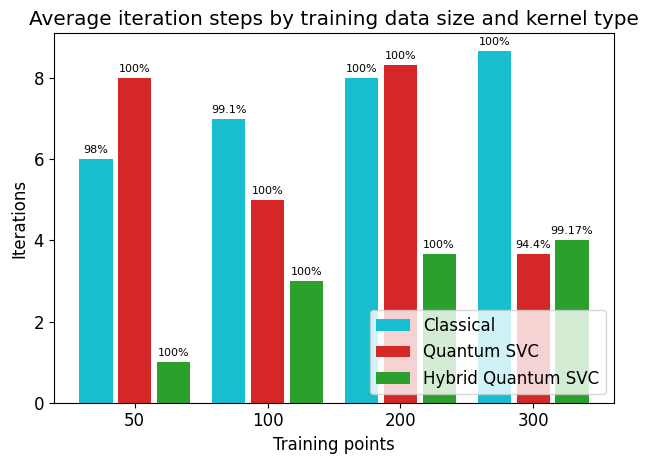}}
    \caption{Illustration of the average experiment results for the Qiskit dataset across various seeds, showing training points against iteration counts for each computing method, with accuracy displayed atop each bar, as detailed in Table \ref{table:results}.}
    \label{fig:bar_plot}
\end{figure}

Table~\ref{table:results} shows the results and Figure~\ref{fig:bar_plot} visualises them.
Note that due to computational overhead, we omitted the results for 500 data points. 
The consistency of the triple-hybrid method (depicted as Hybrid QSVM or HQSVM in the figure) is evident in Figure~\ref{fig:bar_plot}.
We see that the number of iterations increases somewhat linearly with the number of training points used.
This linear relationship is independent from the used method. 

In contrast, the quantum SVM method does not exhibit a linear iteration pattern, likely due to the nature of non-intrinsic optimization methods, which incorporate elements of randomness. This behavior is observed in practice across gate-based classification methods~\cite{Neumann2023} and is also supported in theory, where a theoretical minimum number of iterations exists for gradient-free optimization methods in quantum circuits. This minimum is primarily bounded by the Lipschitz constant, depending on noise levels and algorithmic complexity~\cite{Kungurtsev_2024}. Additionally, the QSVM strategy adheres to the maximum iteration limit set by the user, 10 in this case, and selects the best result obtained within that range.

The triple-hybrid quantum method demonstrated the highest average accuracy across varying training and testing points compared to both the classical and purely quantum SVM methods.
The classical method improves in accuracy as the dataset size increases, while both quantum-based methods experience a slight decline in accuracy when applied to larger datasets.
This might be due to statistical inaccuracies or over-fitting. 

\begin{figure}[b]
\centering
    \centerline{\includegraphics[width=1.4\textwidth]{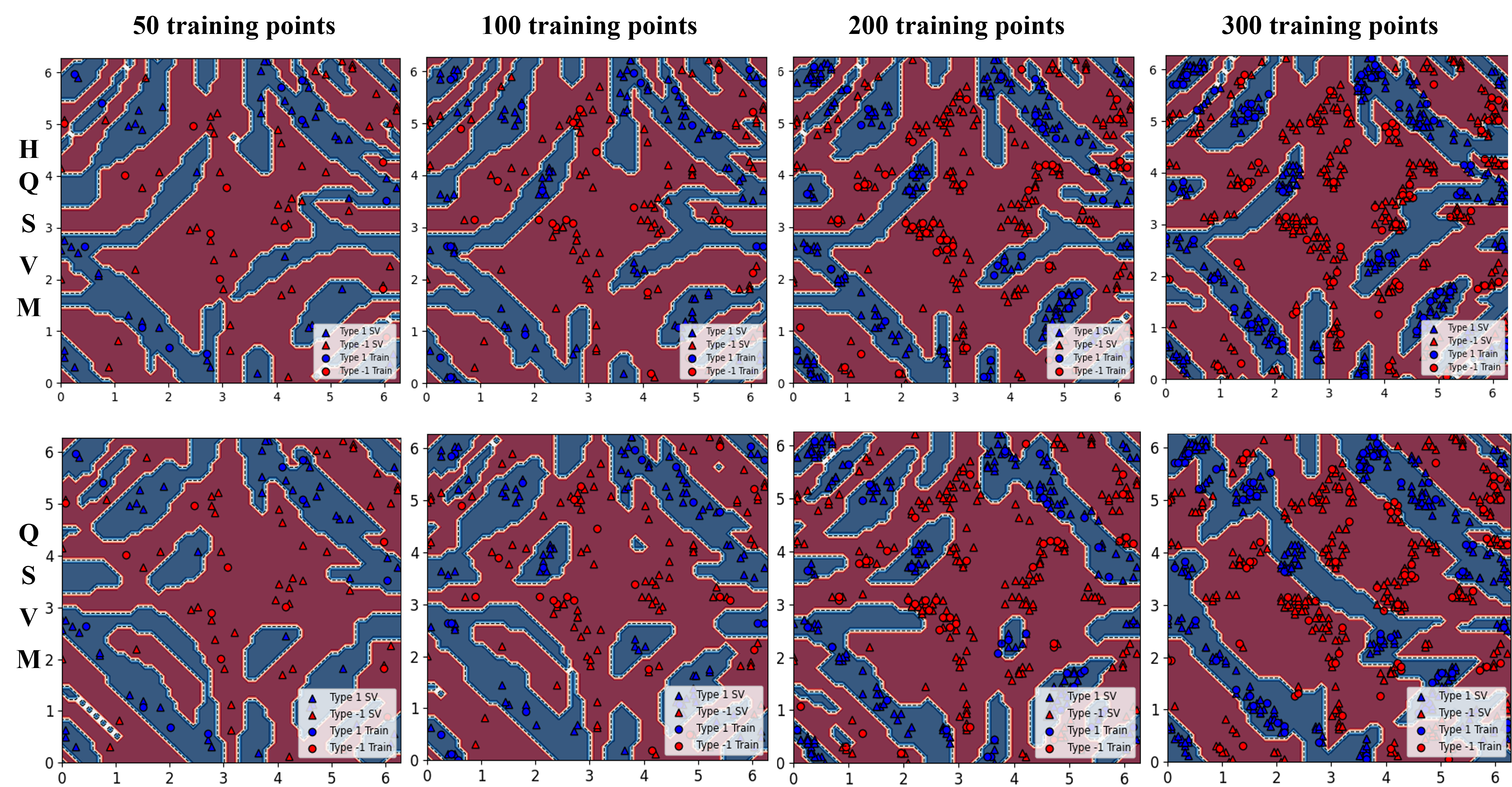}}
    \caption{Classification maps for the qiskit dataset for different training points (seed 600) depicting blue for the -1 label and red for the +1 label are shown for both the QSVM and the hybrid QSVM methods. The circular dots in red and blue represent the training data for labels $+1$ and $-1$, respectively. The triangles in red and blue represent the testing data for labels $+1$ and $-1$, respectively.}
    \label{fig:training_plot}
\end{figure}

The most notable result is however the high expressivity of the quantum methods. 
Figure~\ref{fig:training_plot} shows the entire classification map for the QSVM and the hybrid QSVM (HQSVM) for a varying number of training points. 
Data points with label -1 are shown in blue, and similarly, data points with label +1 are shown in red. 

Empirical observations show that the classification map for the hybrid QSVM method, when trained on a small dataset, closely resembles the map obtained with a larger dataset. 
This implies that with few data points, complex relations can already be found. 
The purely quantum SVM method has difficulty quickly finding the complex landscape. 
The classification maps for few and many training samples found using QSM differ significantly.
Although both methods achieve similar average accuracies across different sizes of training data sets, the hybrid QSVM method appears to be more effective in capturing the classification map with fewer training and testing points.

\begin{figure}[t]
\centering
    \centerline{\includegraphics[width=1\textwidth]{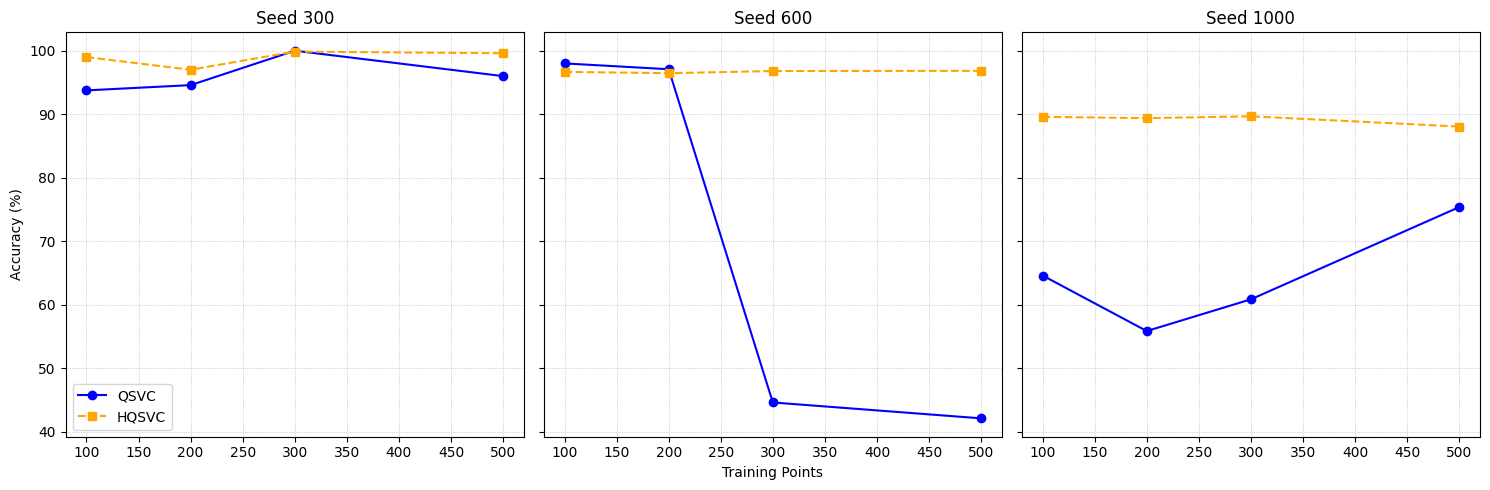}}
    \caption{The accuracy of each method on the qiskit dataset plotted against the varying training data set sizes when predicted using the classification map trained with a size 50 training set for different seeds. The solid lines in blue are the accuracies of the QSVM method while the dotted orange lines are the accuracies of the HQSVM method.}
    \label{fig:extrapolation_plot}
\end{figure}

To observe this claim in a quantifiable way, the varying sizes of the training and testing data sets of each method and seed were plotted on the classification maps created by the smallest training data sets of the respective seeds. The total accuracy of each map and the size of the training point can be seen in Figure~\ref{fig:extrapolation_plot}.

As illustrated in Figure~\ref{fig:extrapolation_plot}, the hybrid QSVM method maintains high accuracy, $\gtrsim 90\%$, when predicting labels for training and testing datasets that are significantly larger than the initial training data for all seeds. This is not the case for the purely quantum method, which only results in a good performance using the data provided in seed 600.
Note that we included 500 data points here for completeness, as only few training iterations were needed. 

\section{Conclusion}
We have introduced a hybrid quantum support vector machine (HQSVM) that integrates quantum gate-based computing, quantum annealing, and classical computation. 
Benchmarking the implementation of HQSVM to a quantum data set showed that it requires fewer iterations on average to accurately classify data points and achieves reliable classification boundaries using fewer training data, thereby greatly outperforming purely quantum methods and classical methods.
Additionally, the HQSVM required fewer training iterations to approximate the entire classification map than the QSVM or classical method. 

However, when executed on two non-quantum, real-world data sets, it did not perform objectively well. For the banknote authentication dataset, it demonstrated better performance than the other methods, yet its overall results were still poor. On the breast cancer dataset, it was shown that the HQSVM solver performed worse than the classical and QSVM solvers. This indicates that two main factors influence the model's accuracy. The first is the structure of the data, likely related to the dataset's periodicity, as the generated periodic dataset yielded strong performance. The second is the inefficiency in transitioning from local to global minima during the annealing phase, since the angles of the parametric variables were determined by $\alpha$ and $\beta$. While speculative interpretations are possible, this remains a compelling area for further research that could offer deeper insights into the role of quantum computing in classification tasks. Moreover, future work could incorporate additional features and larger datasets for training and validation. Since simulations were used, this approach is computationally expensive on classical computers, as simulating quantum circuits scales exponentially with the number of qubits.

It is important to mention that a key limitation of this approach lies in the current lack of seamless integration between the three computational paradigms. Implementing this algorithm on real hardware would require a tightly coupled infrastructure that supports all three technologies to realise any time-based performance gains.
To accommodate for this limitation, we have used simulation for the quantum operations. 
Follow-up work should look into the effect of running these algorithms on quantum hardware. 
These hardware implementations can then also quantify the impact of noise and decoherence present in real quantum systems on the performance of the algorithms. 
As an intermediate step, numerical noise models can already be integrated into the simulations. 


\bibliography{references}


\begin{thebibliography}{48}
\ifx \bisbn   \undefined \def \bisbn  #1{ISBN #1}\fi
\ifx \binits  \undefined \def \binits#1{#1}\fi
\ifx \bauthor  \undefined \def \bauthor#1{#1}\fi
\ifx \batitle  \undefined \def \batitle#1{#1}\fi
\ifx \bjtitle  \undefined \def \bjtitle#1{#1}\fi
\ifx \bvolume  \undefined \def \bvolume#1{\textbf{#1}}\fi
\ifx \byear  \undefined \def \byear#1{#1}\fi
\ifx \bissue  \undefined \def \bissue#1{#1}\fi
\ifx \bfpage  \undefined \def \bfpage#1{#1}\fi
\ifx \blpage  \undefined \def \blpage #1{#1}\fi
\ifx \burl  \undefined \def \burl#1{\textsf{#1}}\fi
\ifx \doiurl  \undefined \def \doiurl#1{\url{https://doi.org/#1}}\fi
\ifx \betal  \undefined \def \betal{\textit{et al.}}\fi
\ifx \binstitute  \undefined \def \binstitute#1{#1}\fi
\ifx \binstitutionaled  \undefined \def \binstitutionaled#1{#1}\fi
\ifx \bctitle  \undefined \def \bctitle#1{#1}\fi
\ifx \beditor  \undefined \def \beditor#1{#1}\fi
\ifx \bpublisher  \undefined \def \bpublisher#1{#1}\fi
\ifx \bbtitle  \undefined \def \bbtitle#1{#1}\fi
\ifx \bedition  \undefined \def \bedition#1{#1}\fi
\ifx \bseriesno  \undefined \def \bseriesno#1{#1}\fi
\ifx \blocation  \undefined \def \blocation#1{#1}\fi
\ifx \bsertitle  \undefined \def \bsertitle#1{#1}\fi
\ifx \bsnm \undefined \def \bsnm#1{#1}\fi
\ifx \bsuffix \undefined \def \bsuffix#1{#1}\fi
\ifx \bparticle \undefined \def \bparticle#1{#1}\fi
\ifx \barticle \undefined \def \barticle#1{#1}\fi
\bibcommenthead
\ifx \bconfdate \undefined \def \bconfdate #1{#1}\fi
\ifx \botherref \undefined \def \botherref #1{#1}\fi
\ifx \url \undefined \def \url#1{\textsf{#1}}\fi
\ifx \bchapter \undefined \def \bchapter#1{#1}\fi
\ifx \bbook \undefined \def \bbook#1{#1}\fi
\ifx \bcomment \undefined \def \bcomment#1{#1}\fi
\ifx \oauthor \undefined \def \oauthor#1{#1}\fi
\ifx \citeauthoryear \undefined \def \citeauthoryear#1{#1}\fi
\ifx \endbibitem  \undefined \def \endbibitem {}\fi
\ifx \bconflocation  \undefined \def \bconflocation#1{#1}\fi
\ifx \arxivurl  \undefined \def \arxivurl#1{\textsf{#1}}\fi
\csname PreBibitemsHook\endcsname

\bibitem[\protect\citeauthoryear{{Markets and Markets}}{2024}]{marketsandmarkets2024quantum}
\begin{botherref}
\oauthor{\bsnm{{Markets and Markets}}}:
Quantum Computing Market Size, Share \& Growth, 2025.
\url{https://www.marketsandmarkets.com/Market-Reports/quantum-computing-market-144888301.html}.
Accessed: 2025-07-25
(2024)
\end{botherref}
\endbibitem

\bibitem[\protect\citeauthoryear{Soller et~al.}{2025}]{McKinsey:2025}
\begin{botherref}
\oauthor{\bsnm{Soller}, \binits{H.}},
\oauthor{\bsnm{Gschwendtner}, \binits{M.}},
\oauthor{\bsnm{Shabani}, \binits{S.}},
\oauthor{\bsnm{Svejstrup}, \binits{W.}}:
{The Year of Quantum: From concept to reality in 2025}.
\url{https://shorturl.at/1uMXQ}.
Accessed: 2025-07-29
(2025)
\end{botherref}
\endbibitem

\bibitem[\protect\citeauthoryear{Salloum et~al.}{2024}]{salloum2024quantum}
\begin{bchapter}
\bauthor{\bsnm{Salloum}, \binits{H.}},
\bauthor{\bsnm{Lukin}, \binits{R.}},
\bauthor{\bsnm{Mazzara}, \binits{M.}}:
\bctitle{{Quantum Computing in Drug Discovery: A Review of Quantum Annealing and Gate-Based Approaches}}.
In: \bbtitle{International Conference on Computational Optimization}
(\byear{2024}).
\burl{https://openreview.net/forum?id=PHRRdOdUya}
\end{bchapter}
\endbibitem

\bibitem[\protect\citeauthoryear{Neumann et~al.}{2023}]{Neumann2023}
\begin{barticle}
\bauthor{\bsnm{Neumann}, \binits{N.M.P.}},
\bauthor{\bsnm{Heer}, \binits{P.B.U.L.}},
\bauthor{\bsnm{Phillipson}, \binits{F.}}:
\batitle{Quantum reinforcement learning}.
\bjtitle{Quantum Information Processing}
\bvolume{22}(\bissue{2}),
\bfpage{125}
(\byear{2023})
\doiurl{10.1007/s11128-023-03867-9}
\end{barticle}
\endbibitem

\bibitem[\protect\citeauthoryear{Liu and Goan}{2022}]{liu2022hybridgatebasedannealingquantum}
\begin{botherref}
\oauthor{\bsnm{Liu}, \binits{C.-Y.}},
\oauthor{\bsnm{Goan}, \binits{H.-S.}}:
{Hybrid Gate-Based and Annealing Quantum Computing for Large-Size Ising Problems}
(2022).
\url{https://arxiv.org/abs/2208.03283}
\end{botherref}
\endbibitem

\bibitem[\protect\citeauthoryear{Zardini}{2024}]{Zardini2024}
\begin{botherref}
\oauthor{\bsnm{Zardini}, \binits{E.}}:
Hybrid classical-quantum algorithms for optimization and machine learning.
PhD thesis,
Universit{\`a} degli studi di Trento
(2024)
\end{botherref}
\endbibitem

\bibitem[\protect\citeauthoryear{Schuld et~al.}{2015}]{Schuld:2015}
\begin{barticle}
\bauthor{\bsnm{Schuld}, \binits{M.}},
\bauthor{\bsnm{Sinayskiy}, \binits{I.}},
\bauthor{\bsnm{Petruccione}, \binits{F.}}:
\batitle{An introduction to quantum machine learning}.
\bjtitle{Contemporary Physics}
\bvolume{56}(\bissue{2}),
\bfpage{172}--\blpage{185}
(\byear{2015})
\doiurl{10.1080/00107514.2014.964942}
\end{barticle}
\endbibitem

\bibitem[\protect\citeauthoryear{Biamonte et~al.}{2017}]{Biamonte2017}
\begin{barticle}
\bauthor{\bsnm{Biamonte}, \binits{J.}},
\bauthor{\bsnm{Wittek}, \binits{P.}},
\bauthor{\bsnm{Pancotti}, \binits{N.}},
\bauthor{\bsnm{Rebentrost}, \binits{P.}},
\bauthor{\bsnm{Wiebe}, \binits{N.}},
\bauthor{\bsnm{Lloyd}, \binits{S.}}:
\batitle{Quantum machine learning}.
\bjtitle{Nature}
\bvolume{549}(\bissue{7671}),
\bfpage{195}--\blpage{202}
(\byear{2017})
\doiurl{10.1038/nature23474}
\end{barticle}
\endbibitem

\bibitem[\protect\citeauthoryear{Schuld and Petruccione}{2018}]{Schuld2018}
\begin{barticle}
\bauthor{\bsnm{Schuld}, \binits{M.}},
\bauthor{\bsnm{Petruccione}, \binits{F.}}:
\batitle{{Supervised Learning with Quantum Computers}}.
\bjtitle{Quantum Science and Technology}
(\byear{2018})
\doiurl{10.1007/978-3-319-96424-9}
\end{barticle}
\endbibitem

\bibitem[\protect\citeauthoryear{Rebentrost et~al.}{2014}]{Rebentrost2014}
\begin{barticle}
\bauthor{\bsnm{Rebentrost}, \binits{P.}},
\bauthor{\bsnm{Mohseni}, \binits{M.}},
\bauthor{\bsnm{Lloyd}, \binits{S.}}:
\batitle{Quantum support vector machine for big data classification}.
\bjtitle{Phys. Rev. Lett.}
\bvolume{113},
\bfpage{130503}
(\byear{2014})
\doiurl{10.1103/PhysRevLett.113.130503}
\end{barticle}
\endbibitem

\bibitem[\protect\citeauthoryear{Schuld and Killoran}{2019}]{Schuld2019}
\begin{barticle}
\bauthor{\bsnm{Schuld}, \binits{M.}},
\bauthor{\bsnm{Killoran}, \binits{N.}}:
\batitle{{Quantum Machine Learning in Feature Hilbert Spaces}}.
\bjtitle{Phys. Rev. Lett.}
\bvolume{122},
\bfpage{040504}
(\byear{2019})
\doiurl{10.1103/PhysRevLett.122.040504}
\end{barticle}
\endbibitem

\bibitem[\protect\citeauthoryear{Havlíček et~al.}{2019}]{Havl_ek_2019}
\begin{barticle}
\bauthor{\bsnm{Havlíček}, \binits{V.}},
\bauthor{\bsnm{Córcoles}, \binits{A.D.}},
\bauthor{\bsnm{Temme}, \binits{K.}},
\bauthor{\bsnm{Harrow}, \binits{A.W.}},
\bauthor{\bsnm{Kandala}, \binits{A.}},
\bauthor{\bsnm{Chow}, \binits{J.M.}},
\bauthor{\bsnm{Gambetta}, \binits{J.M.}}:
\batitle{Supervised learning with quantum-enhanced feature spaces}.
\bjtitle{Nature}
\bvolume{567}(\bissue{7747}),
\bfpage{209}--\blpage{212}
(\byear{2019})
\doiurl{10.1038/s41586-019-0980-2}
\end{barticle}
\endbibitem

\bibitem[\protect\citeauthoryear{van Dam et~al.}{2020}]{vanDam2020}
\begin{botherref}
\oauthor{\bsnm{Dam}, \binits{T.J.}},
\oauthor{\bsnm{Neumann}, \binits{N.M.P.}},
\oauthor{\bsnm{Phillipson}, \binits{F.}},
\oauthor{\bsnm{Berg}, \binits{H.}}:
{Hybrid Helmholtz machines: a gate-based quantum circuit implementation}.
Quantum Inf Process
\textbf{19}(6)
(2020)
\doiurl{10.1007/s11128-020-02660-2}
\end{botherref}
\endbibitem

\bibitem[\protect\citeauthoryear{Meinhardt et~al.}{2020}]{Meinhardt2020}
\begin{bchapter}
\bauthor{\bsnm{Meinhardt}, \binits{N.}},
\bauthor{\bsnm{Neumann}, \binits{N.M.P.}},
\bauthor{\bsnm{Phillipson}, \binits{F.}}:
\bctitle{{Quantum Hopfield Neural Networks: A New Approach and Its Storage Capacity}}.
In: \bbtitle{Computational Science – ICCS 2020},
pp. \bfpage{576}--\blpage{590}.
\bpublisher{Springer}, \blocation{???}
(\byear{2020}).
\doiurl{10.1007/978-3-030-50433-5_44}
\end{bchapter}
\endbibitem

\bibitem[\protect\citeauthoryear{Park et~al.}{2023}]{Park2023}
\begin{botherref}
\oauthor{\bsnm{Park}, \binits{S.}},
\oauthor{\bsnm{Park}, \binits{D.K.}},
\oauthor{\bsnm{Rhee}, \binits{J.-K.K.}}:
Variational quantum approximate support vector machine with inference transfer.
Sci. Rep.
\textbf{13}(1)
(2023)
\doiurl{10.1038/s41598-023-29495-y}
\end{botherref}
\endbibitem

\bibitem[\protect\citeauthoryear{Schnabel and Roth}{2024}]{schnabel2024quantumkernelmethodsscrutiny}
\begin{botherref}
\oauthor{\bsnm{Schnabel}, \binits{J.}},
\oauthor{\bsnm{Roth}, \binits{M.}}:
{Quantum Kernel Methods under Scrutiny: A Benchmarking Study}
(2024).
\url{https://arxiv.org/abs/2409.04406}
\end{botherref}
\endbibitem

\bibitem[\protect\citeauthoryear{Xu et~al.}{2024}]{Xu_2024}
\begin{barticle}
\bauthor{\bsnm{Xu}, \binits{L.}},
\bauthor{\bsnm{Zhang}, \binits{X.-y.}},
\bauthor{\bsnm{Li}, \binits{M.}},
\bauthor{\bsnm{Shen}, \binits{S.-q.}}:
\batitle{Quantum support vector machine for multi classification}.
\bjtitle{Commun. Theor. Phys.}
\bvolume{76}(\bissue{7}),
\bfpage{075105}
(\byear{2024})
\doiurl{10.1088/1572-9494/ad48fc}
\end{barticle}
\endbibitem

\bibitem[\protect\citeauthoryear{Suzuki et~al.}{2024}]{Suzuki2024}
\begin{botherref}
\oauthor{\bsnm{Suzuki}, \binits{T.}},
\oauthor{\bsnm{Hasebe}, \binits{T.}},
\oauthor{\bsnm{Miyazaki}, \binits{T.}}:
Quantum support vector machines for classification and regression on a trapped-ion quantum computer.
Quantum Mach. intell.
\textbf{6}(1)
(2024)
\doiurl{10.1007/s42484-024-00165-0}
\end{botherref}
\endbibitem

\bibitem[\protect\citeauthoryear{Neumann et~al.}{2019}]{Neumann2019}
\begin{barticle}
\bauthor{\bsnm{Neumann}, \binits{N.}},
\bauthor{\bsnm{Phillipson}, \binits{F.}},
\bauthor{\bsnm{Versluis}, \binits{R.}}:
\batitle{Machine learning in the quantum era}.
\bjtitle{Digitale Welt}
\bvolume{3}(\bissue{2}),
\bfpage{24}--\blpage{29}
(\byear{2019})
\doiurl{10.1007/s42354-019-0164-0}
\end{barticle}
\endbibitem

\bibitem[\protect\citeauthoryear{Mansky et~al.}{2023}]{Mansky:2023}
\begin{bchapter}
\bauthor{\bsnm{Mansky}, \binits{M.B.}},
\bauthor{\bsnm{Nüßlein}, \binits{J.}},
\bauthor{\bsnm{Bucher}, \binits{D.}},
\bauthor{\bsnm{Schuman}, \binits{D.}},
\bauthor{\bsnm{Zielinski}, \binits{S.}},
\bauthor{\bsnm{Linnhoff-Popien}, \binits{C.}}:
\bctitle{Sampling problems on a quantum computer}.
In: \bbtitle{2023 IEEE International Conference on Quantum Computing and Engineering (QCE)},
vol. \bseriesno{01},
pp. \bfpage{485}--\blpage{495}
(\byear{2023}).
\doiurl{10.1109/QCE57702.2023.00062}
\end{bchapter}
\endbibitem

\bibitem[\protect\citeauthoryear{Liu et~al.}{2024}]{Liu2024}
\begin{botherref}
\oauthor{\bsnm{Liu}, \binits{J.}},
\oauthor{\bsnm{Liu}, \binits{M.}},
\oauthor{\bsnm{Liu}, \binits{J.-P.}},
\oauthor{\bsnm{Ye}, \binits{Z.}},
\oauthor{\bsnm{Wang}, \binits{Y.}},
\oauthor{\bsnm{Alexeev}, \binits{Y.}},
\oauthor{\bsnm{Eisert}, \binits{J.}},
\oauthor{\bsnm{Jiang}, \binits{L.}}:
Towards provably efficient quantum algorithms for large-scale machine-learning models.
Nature Communications
\textbf{15}(1)
(2024)
\doiurl{10.1038/s41467-023-43957-x}
\end{botherref}
\endbibitem

\bibitem[\protect\citeauthoryear{Rebentrost et~al.}{2018}]{Rebentrost:2018}
\begin{barticle}
\bauthor{\bsnm{Rebentrost}, \binits{P.}},
\bauthor{\bsnm{Bromley}, \binits{T.R.}},
\bauthor{\bsnm{Weedbrook}, \binits{C.}},
\bauthor{\bsnm{Lloyd}, \binits{S.}}:
\batitle{Quantum hopfield neural network}.
\bjtitle{Phys. Rev. A}
\bvolume{98},
\bfpage{042308}
(\byear{2018})
\doiurl{10.1103/PhysRevA.98.042308}
\end{barticle}
\endbibitem

\bibitem[\protect\citeauthoryear{Harrow et~al.}{2009}]{HHL2009}
\begin{barticle}
\bauthor{\bsnm{Harrow}, \binits{A.W.}},
\bauthor{\bsnm{Hassidim}, \binits{A.}},
\bauthor{\bsnm{Lloyd}, \binits{S.}}:
\batitle{Quantum algorithm for linear systems of equations}.
\bjtitle{Phys. Rev. Lett.}
\bvolume{103},
\bfpage{150502}
(\byear{2009})
\doiurl{10.1103/PhysRevLett.103.150502}
\end{barticle}
\endbibitem

\bibitem[\protect\citeauthoryear{Willsch et~al.}{2020}]{Willsch:2020}
\begin{barticle}
\bauthor{\bsnm{Willsch}, \binits{D.}},
\bauthor{\bsnm{Willsch}, \binits{M.}},
\bauthor{\bsnm{{De Raedt}}, \binits{H.}},
\bauthor{\bsnm{Michielsen}, \binits{K.}}:
\batitle{Support vector machines on the d-wave quantum annealer}.
\bjtitle{Computer Physics Communications}
\bvolume{248},
\bfpage{107006}
(\byear{2020})
\doiurl{10.1016/j.cpc.2019.107006}
\end{barticle}
\endbibitem

\bibitem[\protect\citeauthoryear{Phillipson et~al.}{2021}]{Phillipson:2021}
\begin{botherref}
\oauthor{\bsnm{Phillipson}, \binits{F.}},
\oauthor{\bsnm{Wezeman}, \binits{R.S.}},
\oauthor{\bsnm{Chiscop}, \binits{I.}}:
Indoor–outdoor detection in mobile networks using quantum machine learning approaches.
Computers
\textbf{10}(6)
(2021)
\doiurl{10.3390/computers10060071}
\end{botherref}
\endbibitem

\bibitem[\protect\citeauthoryear{Boser et~al.}{1992}]{Boser:1992}
\begin{bchapter}
\bauthor{\bsnm{Boser}, \binits{B.E.}},
\bauthor{\bsnm{Guyon}, \binits{I.M.}},
\bauthor{\bsnm{Vapnik}, \binits{V.N.}}:
\bctitle{A training algorithm for optimal margin classifiers}.
In: \bbtitle{Proceedings of the Fifth Annual Workshop on Computational Learning Theory}.
\bsertitle{COLT '92},
pp. \bfpage{144}--\blpage{152}.
\bpublisher{Association for Computing Machinery},
\blocation{New York, NY, USA}
(\byear{1992}).
\doiurl{10.1145/130385.130401}
\end{bchapter}
\endbibitem

\bibitem[\protect\citeauthoryear{Cohen and Tamir}{2014}]{cohen2014}
\begin{barticle}
\bauthor{\bsnm{Cohen}, \binits{E.}},
\bauthor{\bsnm{Tamir}, \binits{B.}}:
\batitle{{D-Wave and predecessors: From simulated to quantum annealing}}.
\bjtitle{International Journal of Quantum Information}
\bvolume{12}(\bissue{03}),
\bfpage{1430002}
(\byear{2014})
\doiurl{10.1142/S0219749914300022}
\end{barticle}
\endbibitem

\bibitem[\protect\citeauthoryear{Alexandrou et~al.}{2022}]{alexandrou2021}
\begin{bchapter}
\bauthor{\bsnm{Alexandrou}, \binits{C.}},
\bauthor{\bsnm{Funcke}, \binits{L.}},
\bauthor{\bsnm{Hartung}, \binits{T.}},
\bauthor{\bsnm{Jansen}, \binits{K.}},
\bauthor{\bsnm{Kühn}, \binits{S.}},
\bauthor{\bsnm{Polykratis}, \binits{G.}},
\bauthor{\bsnm{Stornati}, \binits{P.}},
\bauthor{\bsnm{Wang}, \binits{X.}},
\bauthor{\bsnm{Weber}, \binits{T.}}:
\bctitle{{Investigating the variance increase of readout error mitigation through classical bit-flip correction on IBM and Rigetti quantum computers}}.
In: \bbtitle{Proceedings of The 38th International Symposium on Lattice Field Theory {\textemdash} PoS(LATTICE2021)},
vol. \bseriesno{396},
p. \bfpage{243}
(\byear{2022}).
\doiurl{10.22323/1.396.0243}
\end{bchapter}
\endbibitem

\bibitem[\protect\citeauthoryear{Olivares-Sánchez et~al.}{2020}]{quantum2020019}
\begin{barticle}
\bauthor{\bsnm{Olivares-Sánchez}, \binits{J.}},
\bauthor{\bsnm{Casanova}, \binits{J.}},
\bauthor{\bsnm{Solano}, \binits{E.}},
\bauthor{\bsnm{Lamata}, \binits{L.}}:
\batitle{{Measurement-Based Adaptation Protocol with Quantum Reinforcement Learning in a Rigetti Quantum Computer}}.
\bjtitle{Quantum Reports}
\bvolume{2}(\bissue{2}),
\bfpage{293}--\blpage{304}
(\byear{2020})
\doiurl{10.3390/quantum2020019}
\end{barticle}
\endbibitem

\bibitem[\protect\citeauthoryear{AbuGhanem}{2025}]{AbuGhanem2025}
\begin{barticle}
\bauthor{\bsnm{AbuGhanem}, \binits{M.}}:
\batitle{{IBM} quantum computers: evolution, performance, and future directions}.
\bjtitle{The Journal of Supercomputing}
\bvolume{81}(\bissue{5}),
\bfpage{687}
(\byear{2025})
\doiurl{10.1007/s11227-025-07047-7}
\end{barticle}
\endbibitem

\bibitem[\protect\citeauthoryear{Kadowaki and Nishimori}{1998}]{KadowakiNishimori:1998}
\begin{barticle}
\bauthor{\bsnm{Kadowaki}, \binits{T.}},
\bauthor{\bsnm{Nishimori}, \binits{H.}}:
\batitle{Quantum annealing in the transverse {Ising} model}.
\bjtitle{Phys. Rev. E}
\bvolume{58},
\bfpage{5355}--\blpage{5363}
(\byear{1998})
\doiurl{10.1103/PhysRevE.58.5355}
\end{barticle}
\endbibitem

\bibitem[\protect\citeauthoryear{Nielsen and Chuang}{2011}]{Nielsen_Chuang_2010}
\begin{bbook}
\bauthor{\bsnm{Nielsen}, \binits{M.A.}},
\bauthor{\bsnm{Chuang}, \binits{I.L.}}:
\bbtitle{{Quantum Computation and Quantum Information: 10th Anniversary Edition}},
\bedition{10th} edn.
\bpublisher{Cambridge University Press},
\blocation{USA}
(\byear{2011}).
\doiurl{10.1017/CBO9780511976667}
\end{bbook}
\endbibitem

\bibitem[\protect\citeauthoryear{Phillipson et~al.}{2023}]{Phillipson2023}
\begin{bchapter}
\bauthor{\bsnm{Phillipson}, \binits{F.}},
\bauthor{\bsnm{Neumann}, \binits{N.}},
\bauthor{\bsnm{Wezeman}, \binits{R.}}:
\bctitle{Classification of hybrid quantum-classical computing}.
In: \bbtitle{Computational Science – ICCS 2023},
pp. \bfpage{18}--\blpage{33}.
\bpublisher{Springer}, \blocation{???}
(\byear{2023}).
\doiurl{10.1007/978-3-031-36030-5_2}
\end{bchapter}
\endbibitem

\bibitem[\protect\citeauthoryear{Jattana}{2024}]{Jattana_2024}
\begin{barticle}
\bauthor{\bsnm{Jattana}, \binits{M.S.}}:
\batitle{Quantum annealer accelerates the variational quantum eigensolver in a triple-hybrid algorithm}.
\bjtitle{Physica Scripta}
\bvolume{99}(\bissue{9}),
\bfpage{095117}
(\byear{2024})
\doiurl{10.1088/1402-4896/ad6aea}
\end{barticle}
\endbibitem

\bibitem[\protect\citeauthoryear{Osaba et~al.}{2025}]{osaba2025solvingdroneroutingproblems}
\begin{botherref}
\oauthor{\bsnm{Osaba}, \binits{E.}},
\oauthor{\bsnm{Miranda-Rodriguez}, \binits{P.}},
\oauthor{\bsnm{Oikonomakis}, \binits{A.}},
\oauthor{\bsnm{Petrič}, \binits{M.}},
\oauthor{\bsnm{Ruiz}, \binits{A.}},
\oauthor{\bsnm{Bock}, \binits{S.}},
\oauthor{\bsnm{Kourtis}, \binits{M.-A.}}:
{Solving Drone Routing Problems with Quantum Computing: A Hybrid Approach Combining Quantum Annealing and Gate-Based Paradigms}
(2025).
\url{https://arxiv.org/abs/2501.18432}
\end{botherref}
\endbibitem

\bibitem[\protect\citeauthoryear{Caleffi et~al.}{2018}]{Caleffi:2018}
\begin{bchapter}
\bauthor{\bsnm{Caleffi}, \binits{M.}},
\bauthor{\bsnm{Cacciapuoti}, \binits{A.S.}},
\bauthor{\bsnm{Bianchi}, \binits{G.}}:
\bctitle{Quantum internet: from communication to distributed computing!}
In: \bbtitle{Proceedings of the 5th ACM International Conference on Nanoscale Computing and Communication}.
\bsertitle{NANOCOM '18}.
\bpublisher{Association for Computing Machinery},
\blocation{New York, NY, USA}
(\byear{2018}).
\doiurl{10.1145/3233188.3233224} .
\burl{https://doi.org/10.1145/3233188.3233224}
\end{bchapter}
\endbibitem

\bibitem[\protect\citeauthoryear{Neumann et~al.}{2020}]{neumann2020imperfect}
\begin{bchapter}
\bauthor{\bsnm{Neumann}, \binits{N.M.P.}},
\bauthor{\bsnm{Houte}, \binits{R.}},
\bauthor{\bsnm{Attema}, \binits{T.}}:
\bctitle{Imperfect distributed quantum phase estimation}.
In: \bbtitle{International Conference on Computational Science},
pp. \bfpage{605}--\blpage{615}
(\byear{2020}).
\bcomment{Springer}
\end{bchapter}
\endbibitem

\bibitem[\protect\citeauthoryear{Caleffi et~al.}{2024}]{Caleffi:2024}
\begin{barticle}
\bauthor{\bsnm{Caleffi}, \binits{M.}},
\bauthor{\bsnm{Amoretti}, \binits{M.}},
\bauthor{\bsnm{Ferrari}, \binits{D.}},
\bauthor{\bsnm{Illiano}, \binits{J.}},
\bauthor{\bsnm{Manzalini}, \binits{A.}},
\bauthor{\bsnm{Cacciapuoti}, \binits{A.S.}}:
\batitle{Distributed quantum computing: A survey}.
\bjtitle{Computer Networks}
\bvolume{254},
\bfpage{110672}
(\byear{2024})
\doiurl{10.1016/j.comnet.2024.110672}
\end{barticle}
\endbibitem

\bibitem[\protect\citeauthoryear{Boschero et~al.}{2025}]{Boschero:2025}
\begin{bchapter}
\bauthor{\bsnm{Boschero}, \binits{J.C.}},
\bauthor{\bsnm{Neumann}, \binits{N.M.P.}},
\bauthor{\bsnm{Schoot}, \binits{W.}},
\bauthor{\bsnm{Phillipson}, \binits{F.}}:
\bctitle{Non-local phase estimation with a rydberg-superconducting qubit hybrid}.
In: \bbtitle{Proceedings of IEEE Symposium on Quantum Software: Quantum Software Engineering}
(\byear{2025}).
\burl{https://arxiv.org/abs/2505.17842}
\end{bchapter}
\endbibitem

\bibitem[\protect\citeauthoryear{Peruzzo et~al.}{2014}]{Peruzzo2014}
\begin{botherref}
\oauthor{\bsnm{Peruzzo}, \binits{A.}},
\oauthor{\bsnm{McClean}, \binits{J.}},
\oauthor{\bsnm{Shadbolt}, \binits{P.}},
\oauthor{\bsnm{Yung}, \binits{M.-H.}},
\oauthor{\bsnm{Zhou}, \binits{X.-Q.}},
\oauthor{\bsnm{Love}, \binits{P.J.}},
\oauthor{\bsnm{Aspuru-Guzik}, \binits{A.}},
\oauthor{\bsnm{O’Brien}, \binits{J.L.}}:
A variational eigenvalue solver on a photonic quantum processor.
Nat Commun
\textbf{5}(1)
(2014)
\doiurl{10.1038/ncomms5213}
\end{botherref}
\endbibitem

\bibitem[\protect\citeauthoryear{Cerezo et~al.}{2021}]{Cerezo2021}
\begin{barticle}
\bauthor{\bsnm{Cerezo}, \binits{M.}},
\bauthor{\bsnm{Arrasmith}, \binits{A.}},
\bauthor{\bsnm{Babbush}, \binits{R.}},
\bauthor{\bsnm{Benjamin}, \binits{S.C.}},
\bauthor{\bsnm{Endo}, \binits{S.}},
\bauthor{\bsnm{Fujii}, \binits{K.}},
\bauthor{\bsnm{McClean}, \binits{J.R.}},
\bauthor{\bsnm{Mitarai}, \binits{K.}},
\bauthor{\bsnm{Yuan}, \binits{X.}},
\bauthor{\bsnm{Cincio}, \binits{L.}},
\bauthor{\bsnm{Coles}, \binits{P.J.}}:
\batitle{Variational quantum algorithms}.
\bjtitle{Nat Rev Phys}
\bvolume{3}(\bissue{9}),
\bfpage{625}--\blpage{644}
(\byear{2021})
\doiurl{10.1038/s42254-021-00348-9}
\end{barticle}
\endbibitem

\bibitem[\protect\citeauthoryear{Karush}{1939}]{Karush:1939}
\begin{botherref}
\oauthor{\bsnm{Karush}, \binits{W.}}:
Minima of functions of several variables with inequalities as side conditions.
Master's thesis,
University of Chicago
(1939)
\end{botherref}
\endbibitem

\bibitem[\protect\citeauthoryear{Kuhn and Tucker}{1951}]{KuhnTucker:1951}
\begin{bchapter}
\bauthor{\bsnm{Kuhn}, \binits{H.W.}},
\bauthor{\bsnm{Tucker}, \binits{A.W.}}:
\bctitle{Nonlinear programming}.
In: \bbtitle{Proceedings of the {S}econd {B}erkeley {S}ymposium on {M}athematical {S}tatistics and {P}robability, 1950},
pp. \bfpage{481}--\blpage{492}.
\bpublisher{University of California Press},
\blocation{Berkeley and Los Angeles}
(\byear{1951}).
\doiurl{10.1007/978-3-0348-0439-4_11}
\end{bchapter}
\endbibitem

\bibitem[\protect\citeauthoryear{Wolberg et~al.}{1993}]{breast_cancer_wisconsin_diagnostic_17}
\begin{botherref}
\oauthor{\bsnm{Wolberg}, \binits{W.}},
\oauthor{\bsnm{Mangasarian}, \binits{O.}},
\oauthor{\bsnm{Street}, \binits{N.}},
\oauthor{\bsnm{Street}, \binits{W.}}:
{Breast Cancer Wisconsin (Diagnostic)}.
UCI Machine Learning Repository.
{DOI}: https://doi.org/10.24432/C5DW2B
(1993)
\end{botherref}
\endbibitem

\bibitem[\protect\citeauthoryear{Lohweg}{2012}]{banknote_authentication_267}
\begin{botherref}
\oauthor{\bsnm{Lohweg}, \binits{V.}}:
{Banknote Authentication}.
UCI Machine Learning Repository.
{DOI}: https://doi.org/10.24432/C55P57
(2012)
\end{botherref}
\endbibitem

\bibitem[\protect\citeauthoryear{Javadi-Abhari et~al.}{2024}]{javadiabhari2024quantumcomputingqiskit}
\begin{botherref}
\oauthor{\bsnm{Javadi-Abhari}, \binits{A.}},
\oauthor{\bsnm{Treinish}, \binits{M.}},
\oauthor{\bsnm{Krsulich}, \binits{K.}},
\oauthor{\bsnm{Wood}, \binits{C.J.}},
\oauthor{\bsnm{Lishman}, \binits{J.}},
\oauthor{\bsnm{Gacon}, \binits{J.}},
\oauthor{\bsnm{Martiel}, \binits{S.}},
\oauthor{\bsnm{Nation}, \binits{P.D.}},
\oauthor{\bsnm{Bishop}, \binits{L.S.}},
\oauthor{\bsnm{Cross}, \binits{A.W.}},
\oauthor{\bsnm{Johnson}, \binits{B.R.}},
\oauthor{\bsnm{Gambetta}, \binits{J.M.}}:
Quantum computing with {Qiskit}
(2024).
\url{https://arxiv.org/abs/2405.08810}
\end{botherref}
\endbibitem

\bibitem[\protect\citeauthoryear{Conn et~al.}{2009}]{conn2009introduction}
\begin{bbook}
\bauthor{\bsnm{Conn}, \binits{A.R.}},
\bauthor{\bsnm{Scheinberg}, \binits{K.}},
\bauthor{\bsnm{Vicente}, \binits{L.N.}}:
\bbtitle{{Introduction to Derivative-Free Optimization}}.
\bpublisher{Society for Industrial and Applied Mathematics},
\blocation{Pennsylvania, United States}
(\byear{2009}).
\doiurl{10.1137/1.9780898718768} .
\burl{https://shorturl.at/mh4pu}
\end{bbook}
\endbibitem

\bibitem[\protect\citeauthoryear{Kungurtsev et~al.}{2024}]{Kungurtsev_2024}
\begin{barticle}
\bauthor{\bsnm{Kungurtsev}, \binits{V.}},
\bauthor{\bsnm{Korpas}, \binits{G.}},
\bauthor{\bsnm{Marecek}, \binits{J.}},
\bauthor{\bsnm{Zhu}, \binits{E.Y.}}:
\batitle{{Iteration Complexity of Variational Quantum Algorithms}}.
\bjtitle{Quantum}
\bvolume{8},
\bfpage{1495}
(\byear{2024})
\doiurl{10.22331/q-2024-10-10-1495}
\end{barticle}
\endbibitem

\end{thebibliography}

\end{document}